# Excess Thermal Energy and Latent Heat in Nanocluster Collisional Growth

Huan Yang[1], Yannis Drossinos[2], Christopher J. Hogan Jr.[1*]

[1]Department of Mechanical Engineering, University of Minnesota, 111 Church St SE, Minneapolis, MN, 55455
[2]European Commission, Joint Research Centre, 21027 Ispra (VA), Italy
[*]To whom correspondence should be addressed: E-mail: hogan108@umn.edu

Nanoclusters can form and grow by nanocluster-monomer (condensation) and nanocluster-nanocluster (coagulation) collisions. During growth, product nanoclusters have elevated thermal energies due to potential and thermal energy exchange following a collision. Even though nanocluster collisional heating may be significant and strongly-size dependent, no prior theory describes such phenomenon. We derive a model to describe the excess thermal energy, the kinetic energy increase of the product cluster, and latent heat, the heat released to the background upon thermalization of the non-equilibrium cluster, of collisional growth. Both quantities are composed of an enthalpic term, related to potential energy minimum differences, and a size-dependent entropic term, which hinges upon heat capacity and energy partitioning. Example calculations using gold nanoclusters demonstrate that collisional heating can be important and strongly size dependent, particularly for reactive collisions involving nanoclusters composed of 14-20 atoms. Excessive latent heat release may have considerable implications in cluster formation and growth.

**Introduction**

The nucleation and growth of condensed phase, nanometer-scale clusters (nanoclusters)[1] from atomic or molecular precursors is critical in a diverse array of systems, including but not limited to vapor phase synthesis nanomaterial reactors,[2,3,4] combustion systems where soot is a byproduct,[5,6,7] and in the ambient atmosphere where new particle formation events occur.[8,9,10,11,12] Continued nanocluster growth invariably leads to nanoparticles, which are a desired product or undesired pollutant and whose properties are dictated by growth dynamics. Modeling the evolution of nanocluster populations is hence of central interest. Both nucleation, i.e., the condensational formation of stable nanoclusters from vapor phase precursors, and nanocluster growth, i.e., coagulation, are thermally driven binary collision phenomena; two reactants collide with one another yielding a single, typically coalesced, product nanocluster. It is such elementary collisional reactions that determine the evolution of a system of forming and growing nanoclusters.

In modeling binary collision reactions, two common assumptions are invoked: (1) nanoclusters are in thermal equilibrium with their surroundings, and (2) nanoclusters are structureless entities with properties calculable from bulk properties. Though there are notable exceptions,[13,14,15,16,17,18] these assumptions are utilized in classical nucleation theory[19,20,21] and modifications of it,[22,23] and are almost universally invoked in population balance models of nanocluster and nanoparticle growth.[24,25] Combined, such assumptions enable prediction of nanocluster population distribution dynamics based upon knowledge of collision rate coefficients and of the equilibrium coefficients for the elementary reactions considered. However, as nucleation and growth models often fail to reflect accurately experimental measurements (e.g., inferred nucleation rates deviate from measurement by a large amount[22,26,27,28,29]), both of these assumptions merit further scrutiny. With regards to the former assumption, nanocluster condensation and coagulation elementary events are not isothermal (isokinetic) when considering the reactants in isolation. Instead, the internal energy distribution of product nanoclusters will deviate from the isothermal distribution (at the system temperature) because of thermal energy (heat) release during collisional growth and coalescence.[30] Nanocluster cooling via collisions[31,32] with inert gas molecules must occur between condensation and coagulation events to bring nanoclusters to an equilibrium internal energy distribution at the system temperature. With regards to the latter assumption, a number of studies[33,34,35,36,37] reveal that nanoclusters composed of a particular number of monomers may be anomalously stable (magic number clusters with anomalously low potential energy minima) or anomalously unstable (anti-magic clusters with high potential energy minima), an effect which is not captured using structureless, bulk-like assumptions. Experiments and numerical studies of magic and anti-magic number clusters reveal that the latent heats of fusion, heat capacities and melting temperatures of nanoclusters can be strongly size dependent,[38,39,40,41,42,43,44] with the addition or removal of one atom drastically changing their properties. Nanocluster thermodynamics and structure are thus linked, and nanoscale structural effects would be expected to determine the extent of heating during collisional growth.

However, a framework to describe heating during condensation and coagulation has not yet been developed which explicitly accounts for size-dependent nanocluster properties. Without such a framework, it is difficult to quantify the importance of nanocluster collisional heating and the role size plays in the heating process, i.e., the validity of the simplifying assumptions commonly made in modeling formation and growth cannot be adequately verified. To this end, we develop a model capable of predicting both the excess thermal energy in collisional growth, which is defined as the mean thermal (kinetic) energy increase of a collisionally formed nanocluster over its mean thermal energy at the system temperature, and the latent heat of collisional growth, which is distinctly defined as the heat which must be transferred to the surroundings for thermal equilibration. The resulting model is general, in that it does not require assumptions on the structure or bulk properties of nanoclusters, and with minor modification it is applicable to any binary collisional growth process in the gas phase. In the following, we describe the model including how thermal to potential energy partitioning ultimately defines both the excess thermal energy and latent heat of collisional growth. We show that both quantities have not only an enthalpic contribution, which is positive, but also an entropic contribution, which can be positive or negative and is non-negligible. The appreciable size-dependent entropic contribution is distinct to nanoclusters; in bulk matter only the enthalpic term is significant and entropic term is small and size independent. Using the developed model, collisional heat release is explored using embedded atom method (EAM)[45] potential gold nanoclusters as a model system. Calculations suggest that the thermal non-equilibrium arising during non-equilibrium growth will be significant in condensed phase nanocluster formation from the vapor phase.

**Results**
**Caloric curves & effective virial theorem**. To examine heating in collisional growth, we first remark that isolated nanoclusters in the gas phase are typically an order of magnitude or smaller in size than the mean free path of the surrounding gas (at atmospheric pressure). It is thus reasonable to assume that between gas molecule encounters, nanoclusters self-equilibrate, i.e., an isolated nanocluster can be considered to exist in a microcanonical ensemble of a total energy $E = K + U$, where $K$ and $U$ are the internal kinetic (thermal) energy and potential energy of the nanocluster, respectively. This total energy is unique for each nanocluster, but the energy distribution of a population of nanoclusters interacting with a bath, i.e., immersed in an inert gas in equilibrium, is sampled from a canonical ensemble (see the derivation of Eq. (5b)). Potential and thermal energy partitioning is typically described by a caloric curve,[46, 47, 48] which is the functional relationship between the thermal energy and the total energy. In prior work, caloric curves revealed a number of unique, size-dependent features of nanoclusters, including anomalous melting temperatures and latent heats of fusion.[43, 44, 46] Here, we define a modified caloric curve (henceforth referred to as caloric curve), containing equivalent information, as the functional dependence of the mean potential energy on the mean thermal energy:

$$\langle U_i \rangle = g(\langle K_i \rangle). \tag{1}$$

The brackets in Eq. (1) denote time-averaged mean quantities, the subscript denotes number of identical atoms in a nanocluster, and $g$ is a function to be determined. We note that the proposed link between $\langle K_i \rangle$ and $\langle U_i \rangle$ is reminiscent of the virial theorem[49] that relates linearly the time-averaged total kinetic energy to the (time-averaged) total potential energy for closed systems of particles interacting pair-wise via a homogeneous potential. While caloric curves can be probed experimentally, it is rather straightforward to examine them using molecular dynamics simulations. Simulations also enable comparison of microcanonical (NVE) and canonical (NVT) caloric curves, which converge to one another for sufficiently large systems.[50]

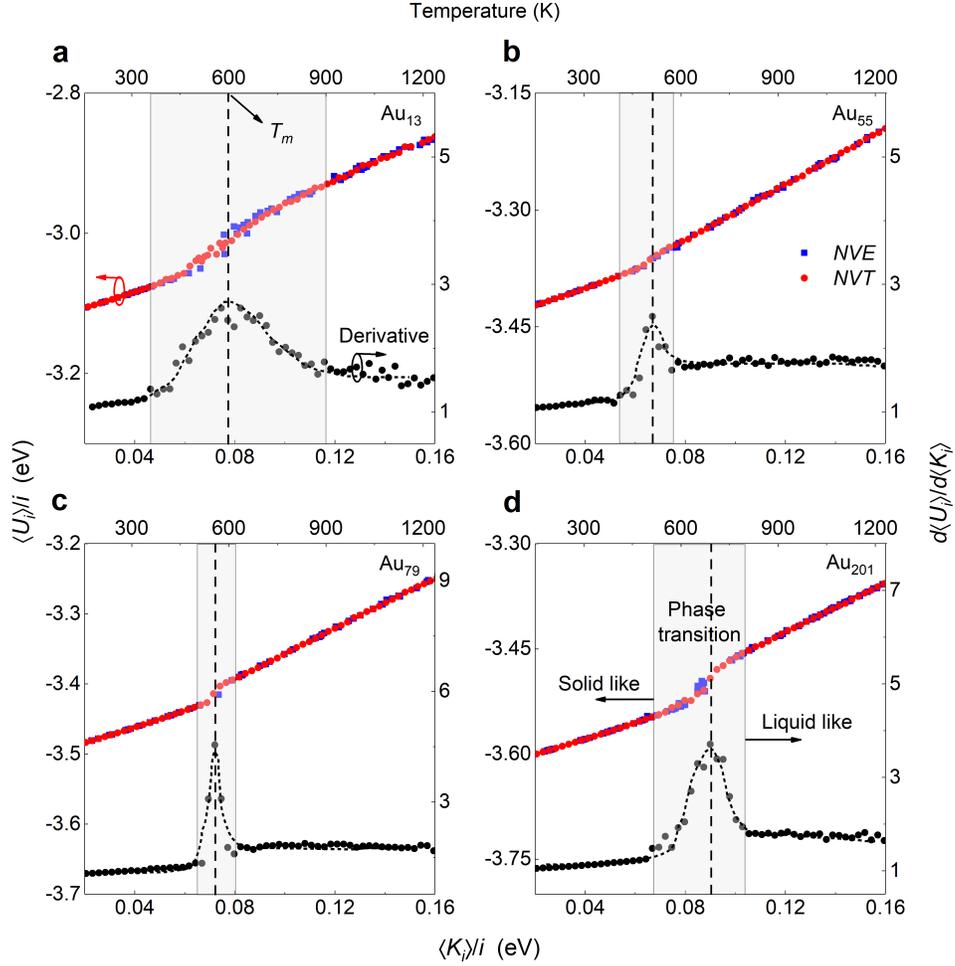

**Fig. 1** Modified caloric curves and their derivatives. **a**, **b**, **c**, **d** Modified caloric curves in NVE (microcanonical, blue squares), NVT (canonical, red circles) ensembles, and their derivatives (black circles) in NVT ensemble are shown for gold nanoclusters composed of 13 (**a**), 55 (**b**), 79 (**c**), and 201 (**d**) atoms. Black dot lines are hand drawing of the trend of derivatives. The modified caloric curves are computed from molecular dynamics simulation using many body EAM (embedded atom method) potential, and their derivatives are obtained from differentiating nearby points on the corresponding modified caloric curve. The upper axis denotes a nanocluster effective temperature scale, defined as $T_{\text{eff}} = 2\langle K_i^T \rangle/(3ik_B)$. The jump in derivative values of modified caloric curves indicates a spread phase transition region which is distinguished from "solid-like" and "liquid-like" region via shading. Peaks of modified caloric curves' derivatives correspond to nanocluster melting temperatures ($T_m$).

Time-averaged potential and thermal energies per atom in the microcanonical and canonical ensembles are calculated (via trajectory simulations) and plotted in Fig. 1 for EAM potential gold nanoclusters composed of 13, 55, 79, and 201 atoms, respectively. Details of the calculations, performed using LAMMPS[51] (Large-scale Atomic/Molecular Massively Parallel Simulator), are provided in Methods. The caloric curves have two nearly linear regions separated by a size-dependent, extended phase-transition region. The phase transition is better identified by plotting $\frac{d\langle U_i \rangle}{d\langle K_i \rangle}$ (also plotted in Fig. 1), a quantity linearly-dependent on the specific heat, which is nearly constant outside the phase change region, and peaks at the point of phase change. We denote the lower energy, constant-slope region of the caloric curve as the "solid-like" phase, and correspondingly the higher energy region as the "liquid-like" phase. Figure S1 (Supplemental Information), displays, and compares to literature results,[52] the inferred melting point of gold nanoclusters as a function of cluster size. We find good agreement in the predicted melting temperature (melting thermal energy) with prior studies. In the regions outside the phase transition region, both microcanonical and canonical curves are well described by the linear relationships

$$\langle U_i \rangle = a_i \langle K_i \rangle + b_i. \qquad (2)$$

Such linear relationships (along with size-dependent nanocluster melting) have been observed experimentally and numerically in previous works.[46, 53] For the liquid-like region of gold nanoclusters with $i = 8 - 1200$, we plot the caloric-curve slope $a_i$ and the normalized potential energy $b_i/i$, the minimum potential energy per atom for the phase in question (i.e., the dissociation energy[54]) in Fig. 2 for both NVE and NVT ensembles. Results are additionally tabulated in Table S1 of the Supplemental Information. As nanocluster size increases, both parameters converge to near constant (bulk) values, $a_1 \to a_\infty$ and $\frac{b_i}{i-1} \to \tilde{b}_\infty < 0$ (the denominator $i - 1$ instead of $i$ will be justified in the discussion of Eq. (6b)). Only small differences are observed between NVE and NVT calculations, which are known to converge to identical caloric curves for sufficiently large clusters.[53, 55] Interestingly, in the $i = 14$ -20 region, both ensembles reveal anomalously large values for the caloric curve slope, and correspondingly low values for the minimum potential energy. The maximum NVE slope is $a_{16}^E = 2.32$. Such nanoclusters, which can be regarded as "magic number" clusters, display unique energy partitioning behavior compared to bulk matter expectations; not only do they have anomalously low minimum energies, but also the increase in mean potential energy per unit increase in thermal energy is anomalously high (compare top left panel to the other panels in Fig. 1).

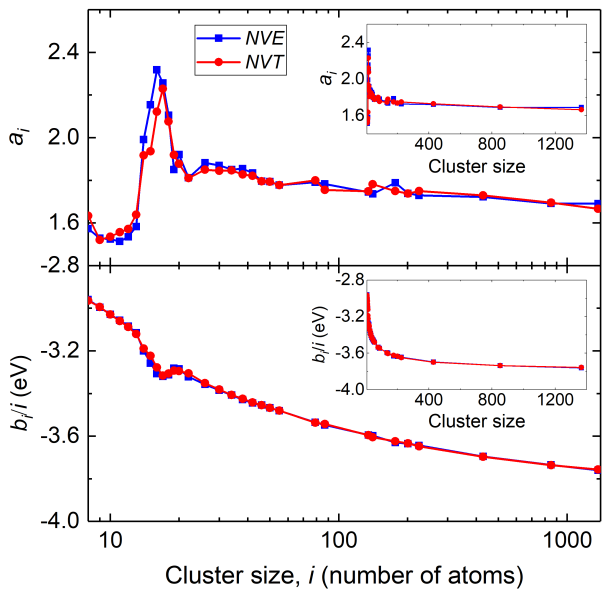

**Fig. 2** Slopes and intercepts of modified caloric curves in the "liquid-like" region for gold. In "liquid-like" regions, modified caloric curves are well described by linear relationships: $\langle U_i \rangle/i = a_i \langle K_i \rangle/i + b_i/i$. The slopes ($a_i$) and intercepts normalized by number of atoms $i$ ($b_i/i$) for EAM simulated gold nanoclusters of varying sizes are shown. NVE denotes simulations in a microcanonical ensemble while NVT denotes simulations in a canonical ensemble. Insets are linear *x*-axis scales. Based on definition, the slopes of modified caloric curve are linearly related to heat capacities of nanoclusters, and the intercepts of modified caloric curves correspond to nanocluster potential energy minima.

**Excess thermal (kinetic) energy of collisional growth**. Changes in caloric-curve slopes and potential energy minima for nanoclusters of different sizes have a direct influence on heat release during nanocluster growth. We study this influence by considering the kinematics of a gas-phase binary reactive collision between two nanoclusters composed of $i$ and $j$ atoms (or a nanocluster and an atom) that approach each other at impact conditions specified by the collision impact parameter $b$. Conservation of energy during collision yields

$$\langle K_i^E \rangle + \langle K_j^E \rangle + \langle U_i^E \rangle + \langle U_j^E \rangle + \frac{1}{2} m_i v_i^2 + \frac{1}{2} m_j v_j^2 + \frac{1}{2} I_i \omega_i^2 + \frac{1}{2} I_j \omega_j^2 =$$
$$\langle K_{i+j}^E \rangle + \langle U_{i+j}^E \rangle + \frac{1}{2} I_{i+j} \omega_{i+j}^2 + \frac{1}{2} m_{i+j} v_{i+j}^2, \quad (3)$$

where $v_i$ is the translational velocity of the center of mass, $\omega_i$ is rotational (angular) velocity with respect to the cluster center of mass, $m_i$ is the nanocluster mass, and $I_i$ is the moment of inertia of an $i$-cluster. The superscript $E$ denotes that the associated quantity was calculated in a NVE simulation. The translational kinetic energy ($\frac{1}{2} m v_i^2$) and rotational energy ($\frac{1}{2} I \omega_i^2$) are explicitly included in Eq. (3) because $\langle K_i^E \rangle$ refers to internal atomic motion relative to the nanocluster center of mass, i.e., it does not include contributions from bulk rotation with respect to the center of mass and center-of-mass translation. We define $c_{ij} = |\boldsymbol{v}_i - \boldsymbol{v}_j|$ the relative speed of reactant nanoclusters and $\mu_{ij} = \frac{m_i m_j}{m_i + m_j}$ their reduced mass. Moreover, the results shown in Fig. 2 suggest that Eq. (1) is applicable to the reactant nanoclusters (which are sampled from a NVT ensemble) and the product nanocluster (which is coagulationally formed in a NVE ensemble as a non-equilibrium activated cluster). Hence, after collision the newly formed product internally self-equilibrates prior to interacting with inert gas molecules. If we substitute Eq. (1) into Eq. (3), neglect reactant nanocluster initial rotation (which is small with respect to the internal degrees of freedom), and use the previously defined quantities, we obtain

$$\langle K_{i+j}^E \rangle + g_{i+j}(\langle K_{i+j}^E \rangle) = \langle K_i^E \rangle + g_i(\langle K_i^E \rangle) + \langle K_j^E \rangle + g_j(\langle K_j^E \rangle) + \frac{1}{2} \mu_{ij} c_{ij}^2 - \frac{1}{2} I_{i+j} \omega_{i+j}^2 \quad .$$
(4a)

Provided $\omega_i$ and $\omega_j$ negligibly influence the rotation of the product, $\omega_{i+j}$ can be approximated from the conservation of angular momentum with respect to the product cluster center of mass, i.e., $\omega_{i+j} = \mu_{ij} c_{ij} b / I_{i+j}$. All the terms on the right hand side of Eq. (4a) are either initially known or can be evaluated from known parameters for a collision, provided $g_i(\langle K_i \rangle)$ is known. If the collision leads to a cluster whose thermal energy does not lie in the phase-transition region, Eq. (2) may be invoked to yield

$$\langle K_{i+j}^E \rangle = \left( \frac{1}{1 + a_{i+j}^E} \right) \Bigg\{ \langle K_i^E \rangle (1 + a_i^E) + \langle K_j^E \rangle (1 + a_j^E) + b_i^E + b_j^E - b_{i+j}^E + \frac{1}{2} \mu_{ij} c_{ij}^2 \Bigg[ 1 -$$
$$\frac{5}{2} \frac{ij}{(i+j)^2} \left( \frac{b}{R_{i+j}} \right)^2 \Bigg] \Bigg\}. \quad (4b)$$

In deriving Eq. (4b), we used the cluster moment of inertia in the rigid rotor approximation, $I_{i+j} = 2(m_i + m_j) R_{i+j}^2$ with $R_{i+j}$ the product nanocluster radius, along with the expression for the cluster angular velocity, and $\mu_{ij} = ij\, m_1 / (i + j)$ since the mass of a cluster composed of identical atoms of mass $m_1$ is $m_{i+j} = (i + j) m_1$.

Equation (4b) shows that the mean thermal energy of the newly formed nanocluster, prior to any inert gas molecule interaction, depends upon: (1) the reacting species mean thermal energies, (2) the difference in potential energy minima between the product and reactants, (3) the slope of the reactant and product caloric curves, which is related to their specific heat, and (4) the translational and rotational energies of the product cluster.

We examine the accuracy of Eq. (4b), specifically the validity of the (approximate) product rotational energy term, using molecular dynamics trajectory calculations[56] of collisional reactions of nanoclusters of $i = 20$ and $j = 34$, with given initial impact parameters and translational velocities. Details are provided in Methods; for simplicity the reactant rotational energies were set to zero. We elected to use unphysically large collision velocities in excess of $10^2$ m s$^{-1}$; the mean thermal speed of an Au$_{20}$ nanocluster would not exceed 120 m s$^{-1}$ until the gas temperature approaches 3000 K, approaching the boiling point of gold. Such large translational velocities were intentionally selected to examine the extent to which the expression for the product rotational energy term is valid, as the translational energy conversion to other degrees of freedom is large in the tested instances. As we show subsequently, in more realistic circumstances, it is small in comparison to the potential and thermal energy terms. The MD simulated thermal energy, calculated as the sum of mean kinetic energies of all atoms in the product nanocluster minus the last term of Eq. (4b), i.e., minus the center of mass translational kinetic energy and rotational energy (equivalently, without the two last terms in Eq. (3)), versus Eq. (4b) predictions is shown in Fig. 3. At small impact parameters and low velocities, mean thermal energies from molecular dynamics collision simulations and from Eq. (4b) are in good agreement. Deviations manifest themselves for large impact parameters and higher velocities, suggesting that in these circumstances the product nanocluster rotational energy is larger in MD simulations than that based on Eq. (4b). We attribute this to use of the rigid body approximation for the product rotational energy. However, instances where disagreement is observed occur for extremely rare high velocity collisions. Overall we find good support for the use Eq. (4b) to calculate the product nanocluster mean thermal energy.

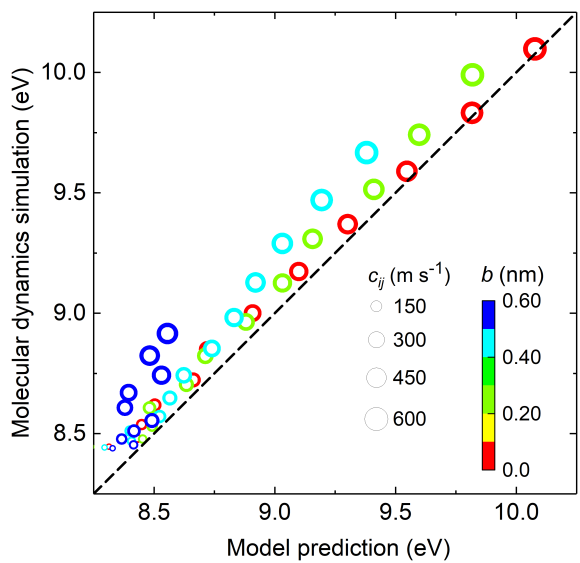

**Fig. 3** Product nanocluster thermal energies. Comparison of the product nanocluster thermal energies from model prediction of Eq. (4b) to molecular dynamics trajectory calculations of the collision between an Au$_{20}$ nanocluster and an Au$_{34}$ nanocluster. The two reactant clusters have zero initial rotational energy with $\langle K_{20}^E \rangle = 2.71\ eV$ and $\langle K_{34}^E \rangle = 3.35\ eV$. Deviations manifest for large impact parameters and higher velocities, suggesting that the product nanocluster rotational energy is larger in molecular dynamics simulations than that based on the model prediction. This is attributed to the use of the rigid body approximation for the product rotational energy.

Further demonstrating the non-equilibrium nature of the product nanoclusters we contrast in Fig. S2 (Supplemental Information) the total thermal energy distribution of a product non-equilibrium nanocluster (in a NVE ensemble) to the thermal energy distribution of an equilibrated nanocluster with the same mean thermal energy, but sampled in a NVT ensemble. Differences in the distribution widths, and hence fluctuation-based (variance defined) specific heats, are apparent. While noteworthy, we show subsequently that only the mean thermal energy is required to assess the extent of nanocluster collisional heating.

We define the excess thermal energy of a product cluster $\Delta K_{ij} = \overline{\langle K_{i+j}^E \rangle} - \langle K_{i+j}^T \rangle$, where $\overline{\langle K_{i+j}^E \rangle}$ is the mean product nanocluster thermal energy, averaged both in time and over all collision possibilities and evaluated by a NVE simulation, and $\langle K_{i+j}^T \rangle$ is the time-averaged thermal energy of the product nanocluster after equilibration with the background gas at temperature $T$ (the superscript $T$ denotes equilibrium parameters in a NVT ensemble). $\overline{\langle K_{i+j}^E \rangle}$ is calculated through the average

$$\overline{\langle K_{i+j}^E \rangle} = \frac{\int_0^\infty \int_0^\infty \int_0^\infty \int_0^{R_\sigma(c_{ij})} \langle K_{i+j}^E \rangle f(E_i^T) f(E_j^T) 2\pi b c_{ij} f(c_{ij}) db\, dc_{ij} dE_i^T dE_j^T}{\int_0^\infty \int_0^{R_\sigma(C)} 2\pi b c_{ij} f(c_{ij}) db\, dc_{ij}}, \quad (5a)$$

where $E_i^T$ is the instantaneous total energy of reactant $i$ with $f(E_i^T)$ its corresponding distribution in equilibrium with the background gas (i.e., a NVT distribution). $f(c_{ij})$ is the Maxwell-Boltzmann speed distribution, and the collision radius is $R_\sigma(c_{ij}) = \left[\frac{\sigma(c_{ij})}{\pi}\right]^{1/2}$, where $\sigma(c_{ij})$ is the reaction cross section for reactant nanoclusters with approaching speed $c_{ij}$. To evaluate the right hand side of Eq. (5a), a link between the time-averaged product thermal energy $\langle K_{i+j}^E \rangle$ and the instantaneous reactant total energies ($E_i^T$ and $E_j^T$) is necessary. This can be established by noting that during close approach, the reactants do not interact with background gas molecules; therefore, their total energies are conserved. This leads to $E_i^T = \langle E_i^E \rangle = \langle K_i^E \rangle + \langle U_i^E \rangle = \langle K_i^E \rangle(1 + a_i^E) + b_i^E$ for each reactant nanocluster: hence, the NVE average in Eq. (4b) may be replaced by the instantaneous, pre-collisional energy $E_i^T$. Moreover, by invoking the ergodic hypothesis $\int_0^\infty E_i^T f(E_i^T) dE_i^T = \langle E_i^T \rangle = \langle K_i^T \rangle + \langle U_i^T \rangle = \langle K_i^T \rangle(1 + a_i^T) + b_i^T$, the approximation $R_\sigma = R_{i+j}$, and Eq. (4b), Eq. (5a) can be written as

$$\overline{\langle K_{i+j}^E \rangle} = \left(\frac{1}{1+a_{i+j}^E}\right)\left\{\langle K_i^T \rangle(1 + a_i^T) + \langle K_j^T \rangle(1 + a_j^T) + b_i^T + b_j^T - b_{i+j}^E + 2k_B T\left[1 - \frac{5ij}{4(i+j)^2}\right]\right\}. \quad (5b)$$

In Eq. (5b), $a_i^T$ and $b_i^T$ denote the slope and intercept of the caloric curve calculated in a NVT simulation, whereas $b_{i+j}^E$ the intercept (of the non-equilibrium product cluster) caloric curve in a NVE simulation. The excess thermal energy of collisional growth is then expressed as

$$\Delta K_{ij} \equiv \overline{\langle K_{i+j}^E \rangle} - \langle K_{i+j}^T \rangle = \left(\frac{1}{1+a_{i+j}^E}\right)\left\{\langle K_i^T \rangle(1 + a_i^T) + \langle K_j^T \rangle(1 + a_j^T) + b_i^T + b_j^T - b_{i+j}^E + 2k_B T\left[1 - \frac{5ij}{4(i+j)^2}\right]\right\} - \langle K_{i+j}^T \rangle. \quad (6a)$$

As mentioned earlier, the thermal energy average $\langle K_i^T \rangle$ does not include the center-of-mass translational degrees of freedom: hence according to the equipartition theorem, or invoking the kinetic theory relationship[50], we define an effective kinetic temperature via $\langle K_i^T \rangle = \frac{3}{2}(i-1)k_B T$. Therefore, Eq. (6a) becomes

$$\Delta K_{ij} = \frac{b_i^T + b_j^T - b_{i+j}^E}{1 + a_{i+j}^E} - T \frac{3}{2} \frac{k_B}{1 + a_{i+j}^E} \left[ i(a_{i+j}^E - a_i^T) + j(a_{i+j}^E - a_j^T) + a_i^T + a_j^T - a_{i+j}^E - \frac{1}{3} + \frac{5ij}{3(i+j)^2} \right].$$
(6b)

Equation (6) is written in the form $\Delta K = \Delta H - T\Delta S$, showing that the excess thermal energy of collision has both enthalpic and entropic contributions. In the large cluster limit, $a_{i+j}^E = a_i^T = a_\infty$, the excess thermal energy of coagulational growth is almost completely determined by enthalpic effects as the entropic contribution is $\frac{3k_B}{2}\left(\frac{1/12 + 3a_\infty}{1 + a_\infty}\right) < \frac{3k_B}{2}$; thus, it is small, cluster-size independent, and bounded by the thermal energy of a single atom. In obtaining the left-hand side of the upper bound of the entropic contribution we used $(i+j)^2 \geq 4ij$. The enthalpic contribution tends to $-\tilde{b}_\infty/(1 + a_\infty)$; the factor $i - 1$ (instead of $i$) is required because otherwise the leading order $\tilde{b}_\infty$ term would cancel out.

Conversely, for nanoclusters, where $a_i^E$ and $a_i^T$ vary with size, the excess thermal energy is also affected by changes in configurational entropy (size dependent), expressed in terms of reactant-to-product changes in the slopes of caloric curves. The entropy is typically positive when $a_{i+j}^E > a_i^T$, and $a_{i+j}^E > a_j^T$, thereby reducing $\Delta K_{ij}$. The reverse applies for $a_{i+j}^E < a_i^T$, and $a_{i+j}^E < a_j^T$. Example calculations using Eq. (6b) are shown in Fig. 4, which displays plots (upper: specific $i$-$j$ pairs, lower: contour plots) of $\Delta K_{ij}$ normalized by $\frac{3}{2}(i+j)k_B$ at system temperatures of 800 K, 1200 K, and 1600K. Linear interpolation is used to calculate the slopes and energy minima which were not explicitly calculated using molecular dynamics simulations. First apparent in Fig. 4 is the magnitude of the excess thermal energy. Nanocluster collisions lead to heating and effective temperature increases of several hundred Kelvin, which is not insignificant in comparison to system temperatures and of relevance because of the strong temperature dependence of dissociation rates. Second, the maximum excess energy is typically found for collisions between equal-sized nanoclusters and the minimum excess energy arises in atom-nanocluster condensation events. In population balance modeling, heat release due to condensation or reaction has been considered previously, but heat release due to coagulation has not; our results show that the excess thermal energy of coagulation is even larger than that of condensation. Third, at lower temperatures, collisions involving low potential energy minimum nanocluster reactants (e.g., Au$_{16}$ and the $i = 14 - 20$ nanoclusters in general) lead to anomalously low excess thermal energies (in comparison to reactions without such nanoclusters). As the system temperature increases, reactive collisions involving these stable reactant nanoclusters have an entropically driven increase in the excess thermal energy of collisional growth, as they have elevated $a_i^T$ values. The opposite is true for reactive collisions where these nanoclusters are products; at lower system temperatures, the excess thermal energies of these reactions are elevated, and as temperature increases, entropic effects reduce their excess thermal energies. The net result of these phenomena is the removal of the "peaks" and "valleys" in plots of the excess thermal energy versus cluster size as system temperature increases.

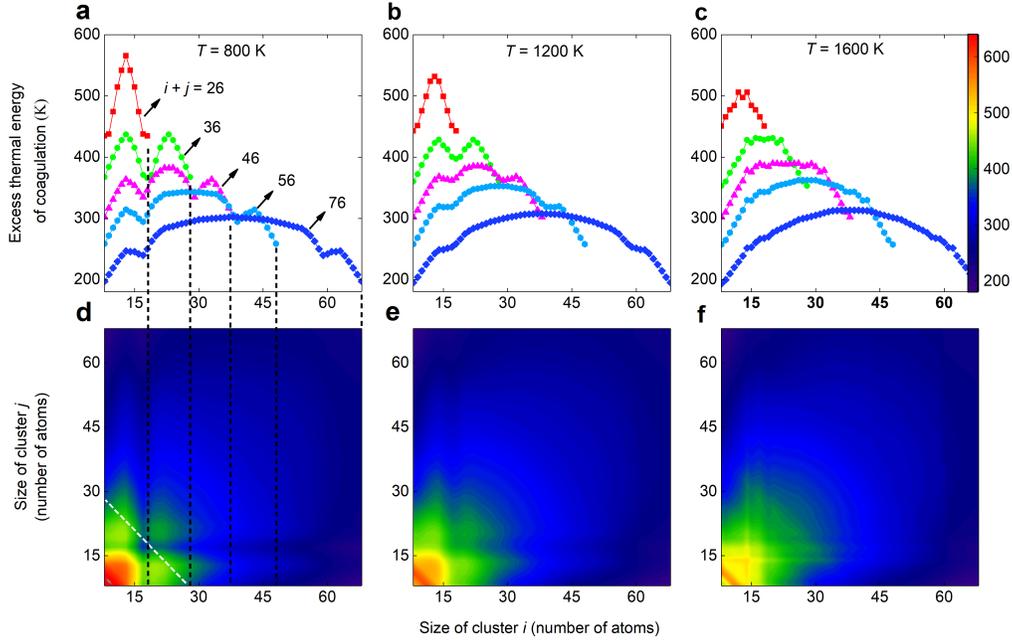

**Fig. 4** Excess thermal energy of coagulation. **a**, **b**, **c** Plots of the excess thermal energy (normalized by $\frac{3}{2}(i+j)k_B$) of coagulation at system temperatures of 800 K (**a**), 1200 K (**b**), and 1600 K (**c**) for selected product gold nanocluster sizes based on the prediction of Eq. (6b). **d**, **e**, **f** Contour plots of the excess energy in terms of reactant gold nanocluster sizes corresponding to the same system temperatures. The white dashed line in (**d**) is an example of the location in the contour plot for the $i+j=36$ line in (**a**). The black vertical dashed lines refer to the maximum size of cluster $i$ (18, 28, 38, 48, and 68) in corresponding curves. For the plotted size range, the magnitude of excess thermal energies is comparable to system temperatures, indicating non-negligible isothermal effects during nanocluster collisional growth.

**Latent heat of collisional growth**. Though related, the latent heat of collisional growth ($L$) is distinct from the excess thermal energy; latent heat is the total energy transferred to the background to equilibrate the product nanocluster at the system temperature. To calculate it from $\Delta K_{ij}$, it is necessary to account for the potential energy to thermal energy conversion occurring within a nanocluster as it is cooled by the background gas. Specifically, the latent heat, defined as the difference between the total energy of the non-equilibrated product cluster and its corresponding thermalized energy, is

$$L \equiv \overline{\langle E_{i+j}^E \rangle} - \langle E_{i+j}^T \rangle = \overline{\langle K_{i+j}^E \rangle} + \overline{\langle U_{i+j}^E \rangle} - \langle K_{i+j}^T \rangle - \langle U_{i+j}^T \rangle. \tag{7a}$$

Following the same assumptions used in determination of the excess thermal energy, the latent heat is expressed as

$$L = b_i^T + b_j^T - b_{i+j}^T - T\frac{3}{2}k_B\left[i\left(a_{i+j}^T - a_i^T\right) + j\left(a_{i+j}^T - a_j^T\right) + a_i^T + a_j^T - a_{i+j}^T - \frac{1}{3} + \frac{5ij}{3(i+j)^2}\right]. \tag{7b}$$

The latent heat also has both enthalpic and entropic contributions, but it only depends on NVT caloric-curve parameters. It is also worth noting that $E_i^T = \langle U_i^T \rangle + \langle K_i^T \rangle = (1+a_i^T)\langle K_i^T \rangle + b_i^T = \frac{3k_BT}{2}(i-1)(1+a_i^T) + b_i^T = c_i^T T(i-1) + b_i^T$, where $c_i^T = \frac{3k_BT}{2}(1+a_i^T)$ is the specific per atom. Therefore, Eq. (7b) can also be written as:

$$L = b_i^T + b_j^T - b_{i+j}^T - T\left[i\left(c_{i+j}^T - c_i^T\right) + j\left(c_{i+j}^T - c_j^T\right) + c_i^T + c_j^T - c_{i+j}^T - \frac{4}{3}k_B + \frac{5ij}{3(i+j)^2}k_B\right].$$

(7c)

Equation (7c) demonstrates that the latent heat of collision can be determined from the dissociation energies $b_i$ and specific heats of reactant and product nanoclusters which can be both measured and calculated.[54] Fig. 5 shows the latent heat normalized by $\frac{3}{2}k_B$. Results parallel the excess thermal energy curves in terms of the influence of nanocluster size for fixed product nanocluster size; anomalously stable clusters lead to local maxima and minima in curves, and otherwise the latent heat is maximized for equal-sized nanocluster collisions. Normalization by $\frac{3}{2}k_B$ highlights the fact that collisional growth reactions involving larger nanoclusters require a larger amount of heat to be transferred to the background gas (although the effective temperatures of such product clusters are lower, cf. Fig. 4, top panels).

Collectively, Eqs. (7a) and (7b) enable prediction of the thermal energy increase for collisionally formed clusters without the need to directly simulate collision dynamics. Instead, computationally less expensive simulations of nanocluster properties in NVE and NVT ensembles can be used to infer thermal energy changes. We therefore believe this set of equations will find utility in developing improved models of nanocluster growth, particularly in high growth rate environments where monomer and nanocluster concentrations are non-negligible in comparison to inert gas molecule concentrations, as well as in closed systems where the latent heat released via collisions leads to increases in system temperature. Future studies will be required to estimate the extent to which thermal non-equilibrium influences nanocluster growth (e.g., by driving dissociation). Collisional heat release may also contribute to the crystallization of gas-phase synthesized materials at system temperatures well below crystallization requirements, which has been observed in previous studies.[57]

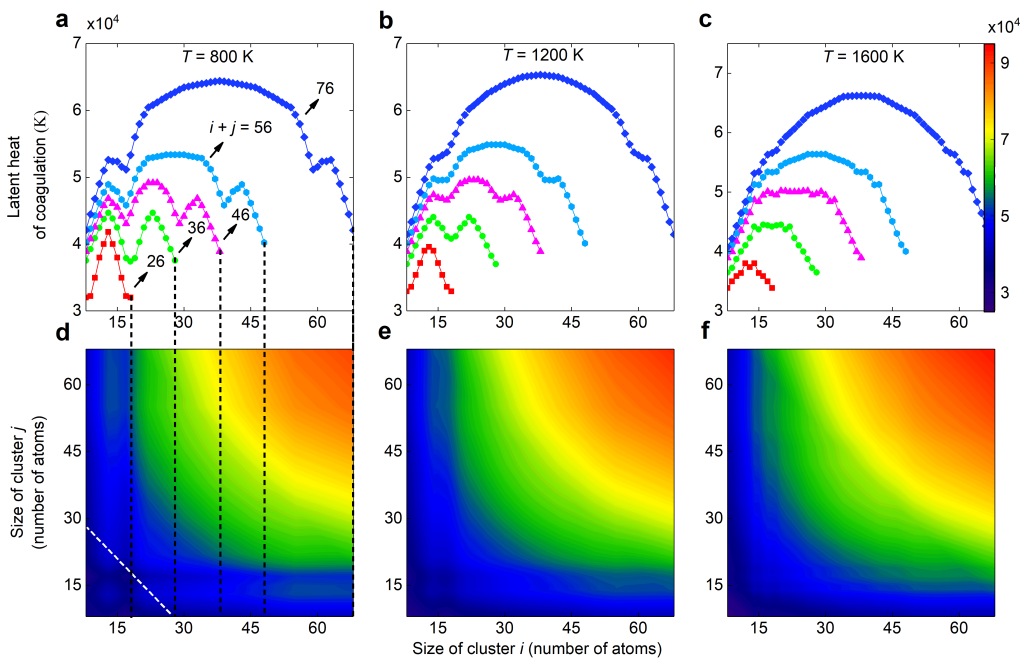

**Fig. 5** Latent heat of coagulation. **a**, **b**, **c** Plots of the latent heat (normalized by $\frac{3}{2}k_B$) of coagulation at system temperatures of 800 K (**a**), 1200 K (**b**), and 1600 K (**c**) based on the prediction of Eq. (7b). **d**, **e**, **f** Contour plots of the latent heat of coagulation in terms of reactant gold nanocluster sizes corresponding to the same system temperatures. The white dashed line in (**d**) is an example of the location in the contour plot for the $i + j = 36$ line in (**a**). The black vertical dashed lines refer to the maximum size of cluster $i$ (18, 28, 38, 48, and 68) in corresponding curves.

As a further matter, these equations additionally afford the opportunity to test the validity of classical nucleation and growth theories. A particularly interesting limit of Eq. (7b) is homogeneous condensation (atom-nanocluster collision), for which $b_1^T = 0$, yielding:

$$L = b_i^T - b_{i+1}^T - T\frac{3}{2}k_B\left[i(a_{i+1}^T - a_i^T) + a_i^T - \frac{1}{3} + \frac{5i}{3(i+1)^2}\right]. \tag{8a}$$

As $i \to \infty$, $-\tilde{b}_\infty + \frac{1}{2}k_BT(1 - 3a_i^T) \to L_\infty(T)$, where $L_\infty(T)$ is the bulk latent heat of vaporization. Wyslouzil & Seinfeld[17] note that according to the capillarity model of nanocluster formation and growth the energy released by a condensing molecule is the bulk latent heat per molecule $L_\infty(T)$ minus the energy required to create the new interface (neglecting a minor $k_BT/2$ term)

$$L = L_\infty(T) - \frac{2}{3}\gamma_\infty A_1 i^{-\frac{1}{3}} + O(i^{-\frac{2}{3}}), \tag{8b}$$

where $A_1$ is the surface area of a monomer calculated from the Van Der Waals radii of gold atom (166 pm), and $\gamma_\infty$ is the surface tension of bulk gold, which is weakly temperature dependent and is taken as 1.1 N m$^{-1}$.[58] The higher order terms in Eq. (8b) arise from size-dependent corrections of the bulk surface tension, from, e.g., the Tolman length correction. $L_\infty(T)$ can be approximated using the latent heat of vaporization of gold $[L_\infty(T_b) = 3.48eV]$ at its boiling point ($T_b = 3243\ K$), and through the large size limit of Eq. (8a), noting that $L_\infty(T) = L_\infty(T_b) + \frac{3k_B}{2}\left(a_\infty^T - \frac{1}{3}\right)(T_b - T)$. Equation (8b) reveals that the classical, structureless cluster model contains no configurational (size dependent) entropic influences. Experimental latent heat of condensation data have been similarly fit to a curve, the size-dependent term monotonically decreasing with cluster size as $i^{-1/4}$, also omitting configurational entropic effects.[30] In Fig. 6 we compare Eq. (8b) predictions to Eq. (8a) at temperatures of 800 K and 1600 K (neglecting the higher order terms). Insets on a linear scale show that the caloric-curve slopes and energy minima inferred here approximately recover the bulk limit for the latent heat of vaporization through the good agreement with the classical theory. Admittedly, the approach to the bulk limit is slower than the prediction of the surface tension model, Eq. (8b). At the same time, calculations show the influence of temperature for nanoclusters with anomalous energy minima. In particular for $i < 25$, unique energy partitioning has large effects on the latent heat of condensation; changes in product cluster size by just several atoms change the latent heat by 16% at low temperature. Such strongly size-dependent, non-monotonic variations suggest that the classical model poorly describes collisional heat release for the smallest nanoclusters.

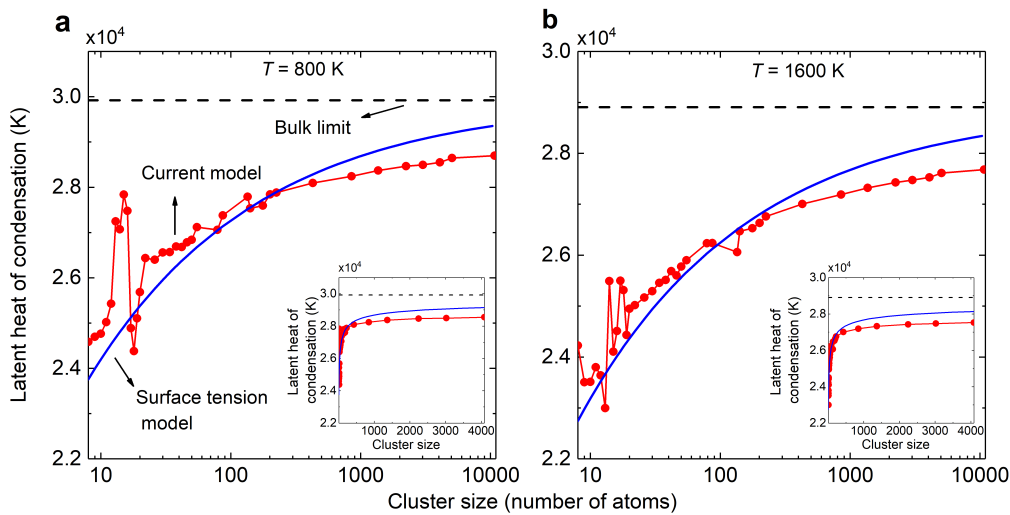

**Fig. 6** Latent heat of condensation. **a**, **b** The latent heat of condensation (normalized by $\frac{3}{2}k_B$) predicted by the capillarity approximation of Classical Nucleation Theory, referred to as surface tension model (Eq. (8b), blue, and Eq. (8a), red), and the bulk latent heat of vaporization for gold at temperatures of 800 K (**a**) and 1600 K (**b**). Insets are linear x-axis scales. Both the current model and surface tension model recovery the bulk limit at large cluster size.

**Discussion**

To summarize, we have developed an approach to calculate the excess thermal energy and latent heat for nanocluster collisional growth reactions. We used it with gold nanoclusters as a case study to show that collisional heating during gas phase growth is significant, with both enthalpic and entropic effects playing a role for nanocluster reactions. Entropic effects are particularly relevant for magic number and anti-magic number clusters, whose anomalous stability gives rise to unique potential-thermal energy partitioning. Implementation of the developed model only requires evaluation of caloric curves in NVE and NVT simulations. Modifications of the presented equations are necessary to account for phase change, as phase transitions lead to non-linear caloric curves.

The interplay between energy partitioning in NVE and NVT ensembles, which determines the excess thermal energy of a collisionally formed nanocluster, is a unique feature of dispersed gas phase systems, as it is these systems where nanoclusters are uniquely afforded time to self-equilibrate before communicating with a thermal bath. A noted previously,[32] the establishment of non-equilibrium conditions and the relaxation of nanoclusters to equilibrium states would be pressure dependent, with formation and growth rates decreasing with decreasing pressure. This prediction is consistent with some,[59] but not all[60] experimental observations of nanocluster formation and growth. To fully understand and quantify the importance of this phenomenon, subsequent reactions of the product nanoclusters need to be studied, including cooling via bath gas collisions and radiation, the dissociation rate at an elevated non-equilibrium energy distribution, and further reactions nanoclusters may undergo in a variety of environments. In principle, just as the changes in potential and thermal energy partitioning with nanocluster size influence collisional heating, they will affect inert gas heat transfer rates and dissociation rates. Along these lines, the present study serves to define the properties of collisionally formed, non-equilibrium nanoclusters whose lifetimes and reactivities remain to be examined, and we suspect the net result will be a complex, material and size dependent network of reactions which must be monitored to comprehensively describe nanocluster formation and growth.

## Methods

**Modified caloric curve calculations**. All simulations are performed using LAMMPS (Large-scale Atomic/Molecular Massively Parallel Simulator) [51]. We first prepare the gold nanoclusters by extracting a spherical region with prescribed radii of a face centered cubic gold crystal (with lattice constants $α = β = γ = 4.078$ Å). Varying the radii of the spherical region results in clusters composed of a different number of gold atoms. The resultant nanoclusters are then equilibrated in a NVT ensemble controlled by a Nosé-Hoover thermostat with the bath temperature uniformly selected in the 200 K to 1400 K range. After complete equilibration, time-averaged thermal and potential energies are calculated to obtain the caloric curve in the NVT ensemble. The averages are taken after thermalization and a given number (40000) of points are averaged. To calculate the caloric curve in the NVE ensemble, we first equilibrate the nanocluster at a selected bath temperature (still uniformly selected in the 200 K to 1400 K range). Upon complete equilibration (100000 time steps), we extract positions and velocities of all atoms at a specific time and transfer them to a NVE ensemble to calculate time-averaged thermal (kinetic) and potential energy. In all simulations, positions and velocities of all atoms are integrated via the velocity Verlet algorithm with a time step of 1 fs. The interatomic non-local potential is based on the many body embedded-atom method (EAM) [45, 61] with the parametrization of Foiles *et al.* [62] In the EAM potential, the total energy $E_{tot}$ is:

$$E_{tot} = \sum_i F_i(\rho_{h,i}) + \frac{1}{2}\sum_{i,j \neq i} \varphi_{ij}(R_{ij}) \qquad (9)$$

where $F_i$ is the embedding term, $\rho_{h,i}$ the local electron density, and $\phi_{ij}(R_{ij})$ is the pair-wise interaction potential with $R_{ij}$ the distance between atoms *i* and *j*.

**Molecular dynamics trajectory calculations**. Following a molecular dynamics trajectory calculation similar to that proposed elsewhere[56], we perform coagulation simulations to validate the predictions of Eq. (4b). Specifically, we are interested in the validity of the approximate rotational energy term in Eq. (4b). We first equilibrate the two reactant nanoclusters in a NVT ensemble at a selected bath temperature higher than the nanocluster melting temperature to obtain "liquid-like" clusters. After equilibration, the two reactant nanoclusters are placed in a 100×100×100 Å simulation domain with periodic boundary conditions. We place the center of mass of nanocluster *i* at the domain center (0, 0, 0), and the center of mass of nanocluster *j* far from the domain center (50, 0, 0) Å to minimize initial interactions. Both nanoclusters have zero rotation with respect to their center of mass initially. To force a coagulation event, a translational velocity with respect to the center of mass pointing towards the negative *x* axis is imposed on nanocluster *j*. The coagulation simulations are performed in a NVE ensemble, and the reactant and product time-averaged thermal (kinetic) energies are recorded (as previously described). Results obtained for different impact parameters and initial center-of-mass translational velocity of nanocluster *j* are shown in Fig. 3, where they are compared with the prediction of equation (4b).

## Data availability

Data for Fig. 2 are provided in Table S1 of the Supplementary information. All other relevant data are available from the corresponding author upon reasonable request.

**Acknowledgements**

This work was supported by Department of Energy Award No. DE-SC0018202. The views expressed are purely those of the authors and may not in any circumstances be regarded as stating an official position of the European Commission. Part of the work was performed while YD was visiting the Department of Mechanical Engineering, University of Minnesota under the Joint Research Centre Visiting Researcher Program.

**Author contributions**

C.J.H. conceived of the issue of elevated thermal energies in collisional growth. H.Y. carried out all molecular dynamics simulations. H.Y., Y.D., and C.J.H. jointly derived the expressions for the excess thermal energy and latent heat and jointly prepared the manuscript. H.Y. prepared all tables and figures and carried out all necessary calculations.

**Competing interests**. The authors declare no competing interests.

**Supplemental information**

**Table S1.** Slope and intercepts (energy minima) of the caloric curves in NVT and NVE ensembles.

| Cluster size | $a^T$ (NVT) | $b^T$ (NVT, eV) | $a^E$ (NVE) | $b^E$ (NVE, eV) |
|---|---|---|---|---|
| 8 | 1.6336 | -2.9649 | 1.5735 | -2.96013 |
| 9 | 1.5205 | -2.994 | 1.528 | -2.99544 |
| 10 | 1.5356 | -3.0295 | 1.524 | -3.0292 |
| 11 | 1.5562 | -3.0601 | 1.5129 | -3.05673 |
| 12 | 1.5709 | -3.088 | 1.5351 | -3.08408 |
| 13 | 1.6385 | -3.1213 | 1.5828 | -3.11485 |
| 14 | 1.9168 | -3.1894 | 1.991 | -3.20107 |
| 15 | 1.936 | -3.2239 | 2.1544 | -3.25787 |
| 16 | 2.1214 | -3.2777 | 2.3181 | -3.30719 |
| 17 | 2.229 | -3.3166 | 2.2561 | -3.31965 |
| 18 | 2.0762 | -3.3068 | 2.1051 | -3.31256 |
| 19 | 1.9186 | -3.2924 | 1.8497 | -3.28174 |
| 20 | 1.8773 | -3.2945 | 1.9204 | -3.28495 |
| 22 | 1.8105 | -3.3057 | 1.8118 | -3.32209 |
| 26 | 1.8496 | -3.3515 | 1.882 | -3.35777 |
| 30 | 1.8444 | -3.3813 | 1.8696 | -3.385 |
| 34 | 1.847 | -3.407 | 1.8503 | -3.40765 |
| 38 | 1.8276 | -3.4253 | 1.8544 | -3.42895 |
| 42 | 1.8214 | -3.4425 | 1.8346 | -3.44429 |
| 46 | 1.7951 | -3.4545 | 1.7963 | -3.45457 |
| 50 | 1.7936 | -3.4676 | 1.7938 | -3.4676 |
| 55 | 1.7775 | -3.4805 | 1.7776 | -3.48055 |
| 79 | 1.7998 | -3.537 | 1.7905 | -3.53544 |
| 87 | 1.7544 | -3.5435 | 1.7827 | -3.5477 |
| 135 | 1.7485 | -3.5956 | 1.7446 | -3.59467 |
| 141 | 1.7809 | -3.60511 | 1.7358 | -3.5983 |
| 177 | 1.7494 | -3.62452 | 1.7885 | -3.63023 |
| 201 | 1.7372 | -3.63453 | 1.7386 | -3.63517 |
| 225 | 1.7493 | -3.64778 | 1.7287 | -3.6438 |
| 429 | 1.7294 | -3.69744 | 1.7219 | -3.6958 |
| 851 | 1.6949 | -3.73643 | 1.6909 | -3.73525 |
| 1366 | 1.6665 | -3.75666 | 1.6904 | -3.76018 |
| 2243 | 1.6492 | -3.77575 | | |
| 3019 | 1.6425 | -3.78635 | | |
| 4087 | 1.6303 | -3.79471 | | |
| 5067 | 1.6464 | -3.80442 | | |

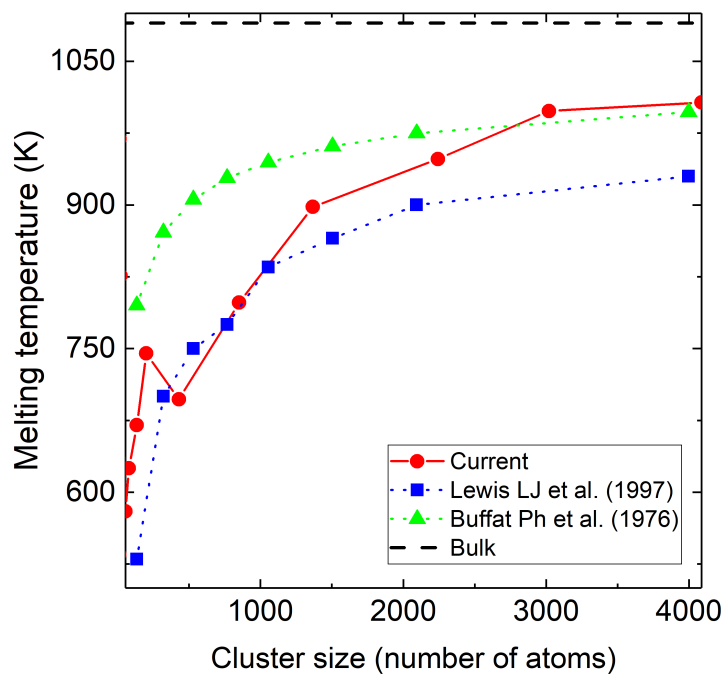

**Fig. S1 Nanocluster melting temperature.** Dependence of gold nanoclusters melting temperature on size are shown (current calculation compared with prior studies[52, 63]). The bulk gold melting temperature is taken as 1090 K, calculated by Lewis *et al.*[52] using the gold EAM potential.

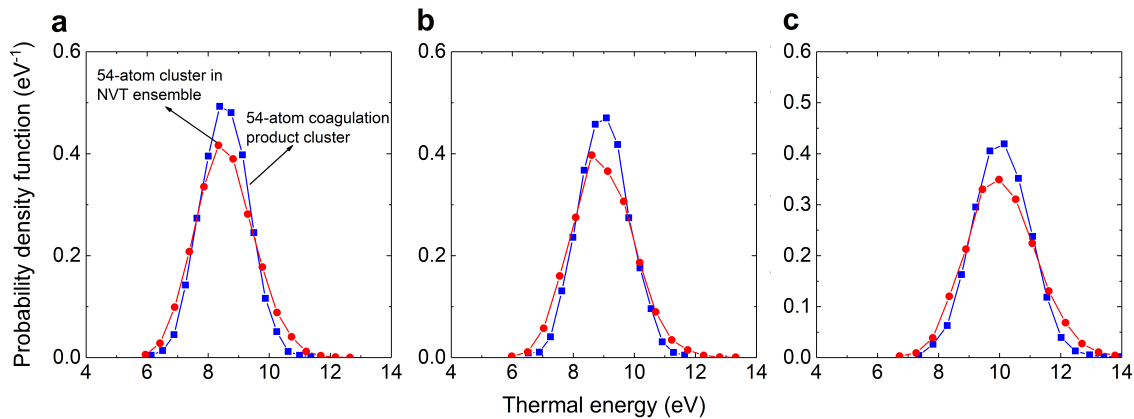

**Fig. S2** Product nanocluster thermal energy distributions. **a**, **b**, **c** Comparison of thermal energy distributions of the 54-atom coagulation product (in a NVE ensemble) and the 54-atom cluster in a NVT ensemble with the same mean thermal energy as the NVE energy. The 54-atom coagulation product is produced by colliding a translating 24-atom cluster and a motionless 30-atom cluster in the NVE ensemble, with $b = 0$, $v = 200$ m s$^{-1}$ (**a**), $b = 0$ $v = 400$ m s$^{-1}$ (**b**), and $b = 0$, $v = 600$ m s$^{-1}$ (**c**), leading to a mean thermal energy of the 54-atom coagulation product of 8.55 eV (**a**), 9.01 eV (**b**), and 10.10 eV (**c**). The coagulation product (in a NVE ensemble) acquires a narrower thermal energy distribution due to its non-equilibrium nature.